\documentclass[aps,prc,groupedaddress,preprintnumbers,
amsmath,amssymb,showpacs]{revtex4}
 
\usepackage{graphicx}
\usepackage{dcolumn}
\usepackage{bm}
\usepackage{epsfig}

\newcommand{\bra}[1]{\mbox{$\langle #1|$}}

\newcommand{\ket}[1]{\mbox{$|#1\rangle$}}

\newcommand{\braket}[2]{\mbox{$\langle #1 | #2 \rangle$}}
\newcommand{\eqn}[1]{\begin{equation} #1 \end{equation}}
\newcommand{\eqna}[1]{\begin{eqnarray} #1 \end{eqnarray}}
\newcommand{\dstyle}[1]{$\ensuremath{\displaystyle{#1}}$}

\begin{document}

\title{New inversion methods for the Lorentz Integral Transform}

\author{Diego Andreasi$^{1,2}$, Winfried Leidemann$^{1,2}$, Christoph Rei{\ss}$^1$, 
and Michael Schwamb$^3$}

\affiliation{$^1$Dipartimento di Fisica, Universit\`a di Trento, I-38050 Povo, Italy}
 
\affiliation{$^2$Istituto Nazionale di Fisica Nucleare, Gruppo Collegato di Trento, 
I-38050 Povo, Italy}

\affiliation{$^3$Institut f\"ur Kernphysik, Johannes Gutenberg-Universit\"at
Mainz, D-55099 Mainz, Germany}

\date{\today}

\begin{abstract}
The Lorentz Integral Transform approach allows microscopic 
calculations of electromagnetic reaction cross sections without explicit 
knowledge of final state wave functions. The necessary inversion of the 
transform has to be treated with great care, since it constitutes a so-called 
ill-posed problem. In this work new inversion techniques for the Lorentz
Integral Transform are introduced. It is shown that they all contain  
a regularization scheme, which is necessary to overcome the ill-posed problem. 
In addition it is illustrated that the new techniques have a much 
broader range of application than the present standard inversion 
method of the Lorentz Integral Transform.

\end{abstract}


\maketitle

\section{Introduction}
About a decade ago the Lorentz Integral Transform (LIT) method has been 
proposed in order to perform ab initio calculations of electroweak reactions 
with nuclei into the continuum \cite{LIT}. The great advantage of the method lies in 
the fact that a calculation of continuum wave functions is not necessary. Indeed the
LIT approach reduces a continuum state problem to a bound-state like problem and 
thus it is sufficient to use bound-state methods. Hence it is not surprising 
that the LIT method has been applied to  
microscopic calculations of quite a few electroweak cross sections of various
nuclei ranging from A=2-7: inclusive electron scattering (see {\it e.g.} 
\cite{PRL1,ELOT04}), total photoabsorption cross sections (see {\it  e.g.} 
\cite{PRL2,PRL3,Sonia}), exclusive electromagnetic reactions 
\cite{Lapiana,Sofia}, photomeson production \cite{Christoph}, and weak 
processes \cite{Nir}. The list of application shows that the LIT approach has
proven to be an important progress opening
up the possibility to carry out microscopic calculations not only for reactions
of classical few-body systems (deuteron, three-body nuclei), but
also for reactions of more complex nuclei. 

The LIT method proceeds in two steps. First one calculates an integral 
transform of the searched function $R(\omega)$ with a Lorentzian-shape kernel:
\begin{equation}                                                     
\label{eqLIT}
L(\sigma_{\rm R},\sigma_{\rm I}) = \int d\omega {\frac {R(\omega)} {(\omega-\sigma_{\rm R})^2 + 
\sigma_{\rm I}^2}} \,,\,\,\,\,\,\,\,\,\,\, \sigma_{\rm R} \,, \sigma_{\rm I} > 0 \,.
\end{equation}                                                       
Then, in a second step, one inverts
the obtained integral transform. The inversion, however, is in principle an
ill-posed problem and the resulting consequences were already 
discussed in  \cite{FBS}. In order to introduce the reader to the problem
we summarize the main points of that discussion. 
The inversion of 
(\ref{eqLIT}) is unstable with respect to high frequency oscillations $\Omega$.
Adding such a high-frequency term $\Delta_\Omega R(\omega)$ to $R(\omega)$ leads
to an additional $\Delta_\Omega L(\sigma_{\rm R},\sigma_{\rm I})$ in the transform.  With 
growing $\Omega$ on the one hand and a constant non negligible oscillation 
amplitude $\Delta_\Omega R$ on the other hand $\Delta_\Omega L$ becomes 
increasingly small and might become even smaller than the size of errors in the 
calculation. Thus $\Delta_\Omega R$ could not be discriminated. If one reduces 
the error in the calculation one can push the not discriminated $\Delta_\Omega 
R$ to higher and higher $\Omega$. However, even if the excluded frequency
range is physically unimportant, one cannot simply find a solution
of the response by application of the inverse operator, since the unphysical
oscillations cannot be separated from the solution. As further explained in
\cite{FBS} one then has to use a regularization scheme for the inversion
in order to avoid such problems. Therefore a specific inversion method
with a built-in regularization was recommended. In fact this method has been used
in all the above mentioned LIT calculations leading to very safe inversion
results.  On the other hand in all these calculations the various LITs have (i) a 
rather simple structure, where essentially only a single peak of $R(\omega)$ has to be 
resolved or (ii) a more complicated structure, which however could be 
subdivided in a sum of simply structured responses, where the various LITs have been 
inverted separately. Case (ii) was already encountered in \cite{LIT}, where the 
inclusive longitudinal deuteron electron response $R(\omega)$ was calculated at 
a constant momentum transfer. There it was necessary to separate out the Coulomb 
monopole and quadrupole transitions,
which lead to a shoulder of the corresponding $R(\omega)$ at the break-up threshold, 
while the rest of the response shows the typical quasi-elastic peak structure. 
Obviously it would be advantageous to invert the total LIT by just one inversion with 
a method appropriated for more complicated structures. In addition it is not 
guaranteed that one can always make the above subdivision in a sum of simply 
structured responses, {\it e.g.} in case of two rather close peaks in the same multipole 
transition. In such cases the standard inversion method would not be
sufficiently precise. 

The aim of the present paper is twofold. The main purpose is the investigation
of the problem of the inversion of a more structured $R(\omega)$ and to introduce 
alternative inversion methods. As already pointed out the LIT method reduces 
a continuum state problem to a bound-state like problem, which can be solved
with typical bound-state techniques. In recent years many different solutions
for the bound-state problem of nuclei with $A>3$ have been developed as
illustrated for example in a rather recent paper about the $^4$He ground-state 
solution with a realistic nucleon-nucleon potential, where seven different 
theoretical methods were applied \cite{He4}. Many of these methods
could also be used for a LIT calculation and as a second aim we want to
show to potential future practitioners of the LIT method that reliable 
inversion techniques exist even in case of more structured response functions.                                            

The paper is organized as follows. In section II we give a brief outline
of the LIT approach. The standard inversion method and alternative approaches
are described in section III. Results with the various inversion techniques  
are discussed in section VI.

\section{The Lorentz Integral Transform method}
\label{sec:th}

In this section we give a brief outline of the LIT technique discussing 
however the approach for inclusive reactions only. The calculation of exclusive 
reactions is somewhat more complicated, but one proceeds in principle
in a very similar manner \cite{Lapiana,Sofia}.

As pointed out in the introduction the starting point of the LIT method \cite{LIT} 
is the calculation of the integral transform of $R(\omega)$ given in 
eq.~(\ref{eqLIT}).
The function $R$ depends on the internal excitation energy $\omega = E_f-E_0$
of a given particle system and contains information about the transition of 
the system from the ground state $|\Psi_0 \rangle$, with energy $E_0$, to the 
final states $|\Psi_f \rangle$, with energy $E_f$, induced by an external probe.
In case of an inclusive reaction $R(\omega)$ denotes the response function
\begin{equation}
\label{Resp}
R(\omega)=\int d\Psi_f|\langle\Psi_f|\hat O|\Psi_0\rangle|^2 
\delta(E_f-E_0-\omega) 
\,,
\end{equation}
where $\hat O$ is a transition operator which characterizes the specific
process under consideration. The response function and in principle also
the LIT are observable quantities. However, in experiment $R(\omega)$
can be determined more directly, while $L(\sigma_{\rm R},\sigma_{\rm I})$ is only 
accessible via the explicit integration (\ref{eqLIT}) of an experimentally over a
sufficiently large $\omega$-range determined $R(\omega)$. Thus also in a 
theoretical calculation it is preferable to eventually determine $R(\omega)$. 
The key point of the LIT method is nonetheless a direct calculation of 
$L(\sigma_{\rm R},\sigma_{\rm I})$, {\it i.e.} without explicit knowledge of $R(\omega)$. Only 
in a second step  $R$ is obtained from the inversion of the transform. The 
great advantage of the method lies in the fact that the generally very 
complicated calculation of final state wave functions $|\Psi_f\rangle$ can be 
avoided as will be discussed below.  On the contrary  a conventional 
calculation of $R(\omega)$ can only be carried out with the explicit knowledge 
of $|\Psi_f\rangle$.

Given an Hamiltonian $H$ of the particle system under consideration one may use
the completeness of the eigenstates of $H$ to show that $L(\sigma_{\rm R},\sigma_{\rm I})$ 
is determined via the following differential equation \cite{LIT}
\begin{equation}
(H-E_0 - \sigma^*) |\tilde\Psi\rangle = \hat O |\Psi_0\rangle   
\label{DE_incl}
\end{equation}
with
\begin{equation}
\sigma =\sigma_{\rm R} + i \sigma_{\rm I} \,.
\end{equation}
Since $H$ has a real eigenvalue spectrum the corresponding homogeneous 
differential equation has only the trivial solution. Thus the solution of 
(\ref{DE_incl}) is unique. This solution leads directly to the transform,
in fact one finds
\begin{equation}
\label{Lnorm}
L(\sigma_{\rm R},\sigma_{\rm I}) = \langle \tilde\Psi|\tilde\Psi\rangle \,.
\end{equation}
Different from a Schr\"odinger equation at positive energies one has a very 
simple boundary condition for the solution of (\ref{DE_incl}). Due to the 
localized source at the right hand side of (\ref{DE_incl}) and because $\sigma$ 
is complex, $\tilde\Psi$ vanishes at large 
distances similar to a bound-state wave function. Therefore one can apply  
similar techniques as for the calculation of a bound-state wave function. 
It is evident that the much more complicated continuum state problem is 
completely avoided by the LIT method. 

\section{Inversion methods for the Lorentz Integral Transform}

In order to invert the LIT one has to calculate $L(\sigma_{\rm R},\sigma_{\rm I})$ for a 
set of $\sigma_{\rm R}$ values at fixed $\sigma_{\rm I}>0$. If one wants to determine 
$R(\omega)$ in a range $\omega_{\rm min} \le \omega \le \omega_{\rm max}$ one should 
choose $\sigma_{\rm R}$ in the interval $\omega_{\rm min}-\sigma_{\rm I} \le \sigma_{\rm R} \le 
\omega_{\rm max} + \sigma_{\rm I}$. For such $\sigma_{\rm R}$ values the kernel in
(\ref{eqLIT}) takes a resonant form with the width determined by $\sigma_{\rm I}$.
The principal difficulties of the LIT inversion were already discussed in 
the introduction, hence in the following subsections we directly describe
the standard LIT inversion method (subsection A) and the new inversion approaches
(subsections B-D).   

\subsection{Inversion with a set of basis functions}

The present standard LIT inversion method consists in the following ansatz for 
the response function
\begin{equation}
R(\omega') = \sum_{n=1}^N c_n \chi_n(\omega',\alpha_i) \,
\label{sumr}
\end{equation}
with $\omega'=\omega-\omega_{\rm th}$, where $\omega_{\rm th}$ is the threshold energy for the break-up
into the continuum. The $\chi_n$ are given functions with nonlinear parameters $\alpha_i$.
A basis set frequently used in the LIT calculations of [1-11] is
\begin{equation}
\label{bset}
\chi_n(\omega,\alpha_i) = \omega^{\alpha_1} \exp(- {\frac {\alpha_2} {n \omega}}) \,.
\end{equation}
In addition also possible information on narrow levels, like {\it e.g.} elastic transitions,
could be incorporated easily into the set $\chi_n$.

Substituting such an expansion into the right hand side of
(\ref{eqLIT}) one obtains
\begin{equation}
L(\sigma_{\rm R},\sigma_{\rm I}) =
\sum_{n=1}^N c_n \tilde\chi_n(\sigma_{\rm R},\sigma_{\rm I},\alpha_i) \,,
\label{sumphi}
\end{equation}
where
\begin{equation}
\tilde\chi_n(\sigma_{\rm R},\sigma_{\rm I},\alpha_i) = 
\int_0^\infty d\omega' {\frac {\chi_n(\omega',\alpha_i)} {(\omega'-\sigma_{\rm R})^2 + \sigma_{\rm I}^2}} 
\,\,.
\end{equation}

For given $\alpha_i$ the linear parameters $c_n$ are determined from a best fit of 
$L(\sigma_{\rm R},\sigma_{\rm I})$ of eq.~(\ref{sumphi}) to the calculated 
$L(\sigma_{\rm R},\sigma_{\rm I})$ of eq.~(\ref{Lnorm}) for a number of $\sigma_{\rm R}$ points 
much larger than $N$. If one takes the basis set (\ref{bset}) one should vary 
the nonlinear parameter $\alpha_2$ in a rather large range, while one can 
determine $\alpha_1$ from the possibly known threshold behaviour of $R$ or one 
can vary $\alpha_1$ within a reasonable range. 
 
For any values of $N$, $\alpha_1$, and $\alpha_2$ the overall best fit is 
selected and then the procedure is repeated for $N=N+1$ till 
a stability of the inverted response is obtained and taken as inversion 
result. A further increase of $N$ will eventually reach a point, where the 
inversion becomes unstable leading typically to random oscillations. The 
reason is that $L(\sigma_{\rm R},\sigma_{\rm I})$ of eq.~(\ref{sumphi}) is not determined 
precisely enough so that a randomly oscillating $R(\omega)$ leads to a better 
fit than the true response. It is evident that the number of functions $N$ 
plays the role of a regularization parameter and has to be chosen within the 
above mentioned stability region. Normally such a stability region is reached 
without greater problems, however, it can in principle happen that one does not 
find a stable result. Then one can either try to improve the precision of the 
calculated $L(\sigma_{\rm R},\sigma_{\rm I})$ or use different basis sets.  In case that 
the response exhibits an unexpected structure it is useful to decrease the 
parameter $\sigma_{\rm I}$ in order to have a better resolution in the transform.

We should mention that the completeness of the basis set $\chi_n$ is not really
relevant. In fact any basis set is only used up to a relatively low value of $N$. 
Information which can be parametrized only via $\chi_n$ with $n>N$ is lost anyway. 
Therefore it is important to work with a basis set, where a relatively small $N$ 
leads to a high-quality result of $R(\omega)$. Because of the variation of the 
nonlinear parameters $\alpha_i$ of $\chi_n$ many different basis sets are used and 
it is normally no problem to find a proper basis set.

As already pointed out the standard LIT inversion method leads to high-precision
results in case of simply structured response functions, but can have problems in case
of more complicated structures. If one considers however just the low-energy part 
of the response the standard method is very reliable. This is due to the fact that
the break-up threshold $\omega_{\rm th}$ is directly incorporated in the inversion
($R(\omega)$ = 0 for $\omega \le \omega_{\rm th})$ and also a known low-energy behaviour
can be directly implemented ({\it e.g.} by the parameter $\alpha_1$ in (\ref{bset})).
In addition the inversion can be restricted to a small energy interval above 
threshold avoiding problems arising from structures at higher energies
(see also \cite{Diego}).

\subsection{Inversion with wavelets}

\begin{figure}[ht]
\epsfig{file=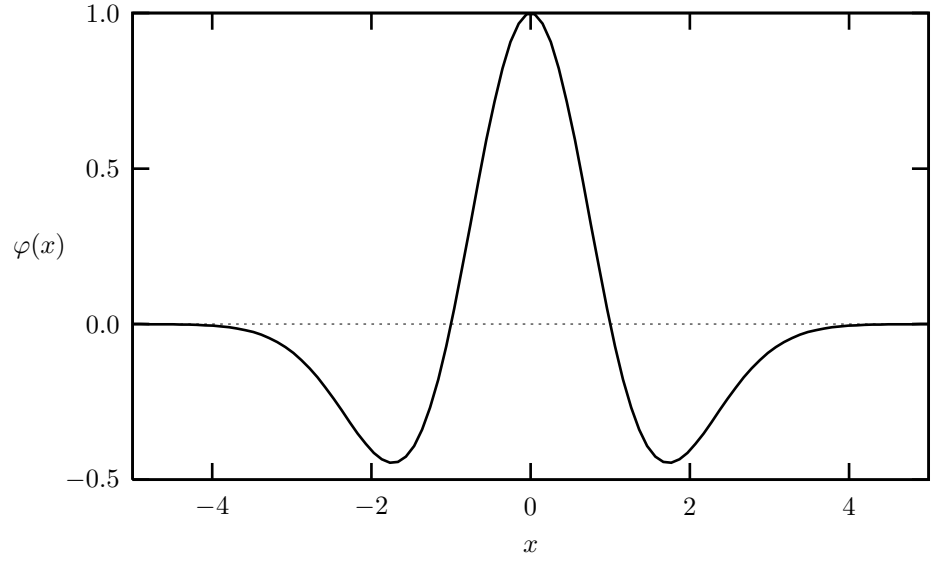}
\caption{\label{fig:01}
  Mexican hat wavelet (eq. (10)).}
\end{figure}

The inversion with wavelets is formally very similar to the inversion with
a set of basis functions described in the preceding subsection.
One starts from the same expansion (\ref{sumr}), but one uses wavelets instead 
of the basis functions $\chi_n$. Here we will take the Mexican hat wavelet
\begin{equation}
\varphi(x)=(1-x^2)~\exp(-{\frac{x^2}{2}}) \,,   
\end{equation}
which is shown in fig.~\ref{fig:01}.
We expand $R(\omega)$ in the range $\omega_{\rm th} \le \omega \le \omega_{\rm max}$ as follows 
\begin{equation}
R(\omega)=\sum_{m=0}^M \sum_{n=1}^{N_m} c_{m,n}\varphi_{m,n}(\omega,\alpha) 
\end{equation}
with
\begin{equation}
\varphi_{m,n}(\omega,\alpha) = \varphi(\alpha \frac{\omega-\omega_{m,n}}{2^m}) \,,
\end{equation}
where $\alpha$ is a nonlinear parameter and the $N_m$ grid points $\omega_{m,n}$ 
are equally distributed ($\omega_{m,n}= n (\omega_{\rm max}-\omega_{\rm th})/(N_m+1))$.
Such an expansion leads to the following LIT
\begin{equation}                                                     
L(\sigma_{\rm R},\sigma_{\rm I}) =                                               
\sum_{m=0}^M \sum_{n=1}^{N_m} c_{m,n} \tilde\varphi_{m,n}(\sigma_{\rm R},\sigma_{\rm I},\alpha)         
\end{equation}                                                       
with 
\begin{equation}                                                     
\tilde\varphi_{m,n}(\sigma_{\rm R},\sigma_{\rm I},\alpha) =                           
\int_{\omega_{\rm th}}^\infty d\omega {\frac {\varphi_{m,n}(\omega,\alpha)} 
        {(\omega-\sigma_{\rm R})^2 + \sigma_{\rm I}^2}} \,\,.    
\end{equation}  
The wavelets $\varphi_{m,n}$ are strongly localized functions about $\omega=\omega_{m,n}$
in contrast to the basis functions $\chi_n$ in subsection A, which generally
are considerably different from zero over a rather large $\omega$-range.
The widths of the wavelets can be modified with the parameter $\alpha$.
In addition with the parameter $M$ one can choose the presence of various scales of 
wavelets. Because of these wavelet characteristics one easily understands 
that they are much more appropriated to represent complicated structure of 
$R(\omega)$ than the basis functions $\chi_n$ of the previous subsection.

For the actual inversion problems discussed in the next section we use up to three
different wavelet scales and various values for the $N_m$ sets (see table~I), in 
addition we include an additional basis function $\varphi_0$, namely a constant 
($\varphi_0=c_0$).
For any of the 25 $N_m$ sets of table~I we proceed as follows. The nonlinear 
parameter $\alpha$ is varied over a wide range and for any value of $\alpha$ a 
best fit is performed leading to the determination of the linear parameters
$c_{m,n}$. For any $N_m$ set the overall best fit is then chosen. Of these
25 inversion results we choose again the overall best fit. In case that
unrealistic oscillations are present in the inversion the next best inversion 
result is taken. Of course such a procedure is nothing else than a regularization
scheme.

\begin{table}
\caption{$M$ and $N_m$ values used for the wavelet inversion of section IV}
\begin{ruledtabular}
\begin{tabular}{cccccccccccc}
      &    &    &    &     &     & $M$=0 & & & & & \\
\colrule
$N_0$ & 15 & 31 & 63 & 127 & 255 & & & & & & \\
\colrule
      &    &    &    &     &     & $M$=1 & & & & & \\
\colrule
$N_0$ & 7 & 15 & 31 & 63 & 127 & 255 & 15 &  31 &  63 & 127 &  255 \\
$N_1$ & 7 & 15 & 31 & 63 & 127 & 255 &  7 &  15 &  31 &  63 &  127 \\
\colrule
      &    &    &    &     &     & $M$=2 & & & & & \\
\colrule
$N_0$ & 7 & 15 & 31 & 63 & 127 &  31 &  63 & 127 & 255 \\
$N_1$ & 7 & 15 & 31 & 63 & 127 &  15 &  31 &  63 & 127 \\
$N_2$ & 7 & 15 & 31 & 63 & 127 &   7 &  15 &  31 &  63 \\  
\end{tabular}      
\end{ruledtabular}                                                             
\end{table}

\subsection{Fridman approach applied to the LIT inversion}

The following method to invert the Lorentz integral transformation is based on the approach
of V. M. Fridman \cite{fridman56}. This method solves by iteration the Fredholm integral 
equation of first order 

\eqn{
  F(x) = \int_a^b\! dy\, K(x,y)\, f(y)\,,
}
which has a solution 
and converges in the Hilbert space of square integrable real functions on $[a,b]$ ($L_2(a,b)$) 
if the kernel $K$ is strictly positive definite and symmetric. 
The iterative equation is given by
\eqn{\label{eqn:fridit}
  f_{l+1}(x) 
  = 
  f_l(x) 
  + \lambda\,
  \left[ F(x) - \int_a^b \! dy\,K(x,y)\, f_l(y)\right]
  \quad,\quad l\in\mathbf{N}_0 
}
with $f_0\in L_2(a,b)$ and $0<\lambda<2\lambda_1$, where $\lambda_1$ is the smallest characteristic 
number of $K$. Using the spectral theorem  it is evident that the iteration of 
eq.~(\ref{eqn:fridit}) diverges if at least one eigenvalue of $K$ is in $]-\infty,0]$.

To apply this method to the inversion of the Lorentz integral transform
\eqn{\label{eqn:lit}
  L(\sigma_{\rm R},\sigma_{\rm I} ) 
  = 
  \int_{\omega_{\rm th}}^\infty \! d\omega K(\sigma_{\rm R},\sigma_{\rm I},\omega) 
  R(\omega)\quad,\quad K(\sigma_{\rm R},\sigma_{\rm I},\omega)
  =
  \frac{1}{(\omega - \sigma_{\rm R})^2+\sigma_{\rm I}^2}\quad,\quad \sigma_{\rm I} >0
}
one has to discretisize the integration and symmetrize the integral kernel $K$.
In practice the integration in eq.~(\ref{eqn:lit}) ranges from $\omega_{\rm th}$ to $\omega_{\rm max}+\sigma_{\rm I}$, 
since contributions beyond $\omega_{\rm max}+\sigma_{\rm I}$ are of minor importance.
The numerical integration is done using the simple formula:
\eqna{\label{eqn:intcr}
  \int_a^b \! dx\, f(x) 
  & = & 
  \sum_{i=0}^n w_i\,f(x_i)\,, \\
  h 
  &=& 
  \frac{b-a}{n}\,,\quad w_0=w_n=\frac{h}{2}\,,\quad w_i=h\,,\quad i\in\{1,\ldots,n-1\}
  \,,
  \nonumber\\
  x_i & = & a+h\,i\,,\quad i\in\{0,\ldots,n\}
  \quad. 
  \nonumber
}  
Therefore the symmetrized iterative equation is given by:
\eqna{\label{eqn:itlit}
  R_{l+1}(x_i) 
  & = & 
  R_{l}(x_i) 
  + \lambda\,\left[ w_i L(x_i,\sigma_{\rm I} ) - \sum_{j=0}^n K_{ij}^{(\sigma_{\rm I} )} R_l(x_j)\right]\,,
  \\\nonumber\\
  K_{ij}^{(\sigma_{\rm I} )}
  & = &
  \frac{w_i w_j}{(x_i-x_j)^2+\sigma_{\rm I}^2}
  \longleftarrow 
  K(x_i,\sigma_{\rm I} ,x_j)\,,\label{eqn:itlitK}
}
where $\sigma_{\rm R} $ and $\omega$ lie on the same grid $(x_i)$. We do not prove that $K_{ij}^{(\sigma_{\rm I} )}$
is strictly positive definite, since the algorithm has to diverge if this is not the case and 
we have to check the convergence numerically anyway.

The iteration is started with the initial function
\eqn{
  R_0(x) = \frac{\sigma_{\rm I} }{\pi}\, L(x,\sigma_{\rm I} )\,,
}
which is motivated by the following definition of the Dirac-Delta-function
\eqn{
  \delta(x) = \lim_{\sigma_{\rm I}\to 0}  \frac{\sigma_{\rm I}}{\pi}\,\frac{1}{x^2+\sigma^2_{\rm I}}\,.
}
Let us remind that extracting $R$ using the limit \dstyle{{\sigma_{\rm I}\to 0}}, {\it i.e.}
\eqn{
  R(\sigma_{\rm R} ) 
  = \lim_{\sigma_{\rm I}\to 0} \frac{\sigma_{\rm I}}{\pi}\,L(\sigma_{\rm R} ,\sigma_{\rm I} ) 
  = \lim_{\sigma_{\rm I}\to 0} \frac{\sigma_{\rm I}}{\pi}\,
  \int_0^\infty \! d\omega\,K(\sigma_{\rm R} ,\sigma_{\rm I} ,\omega)\, R(\omega)\,,
}
is numerically unstable.

The smallest characteristic number of the integral kernel \dstyle{K} is 
the inverse of the largest eigenvalue $\mu_1$ of \dstyle{(K_{ij}^{(\sigma_{\rm I} )})}. 
In this work we have used $\lambda = \frac{1}{\mu_1}$. This value is not optimized. The larger
$\lambda$ is, the faster the algorithm will converge, but the procedure will also become more unstable.

Under real conditions one does not know the solution of the integral equation. One has
to compare the ``input'' Lorentz $L$ with the LIT of the actual result for $R_l$. 
The relative error at every grid point 
\eqn{
  \label{eqn:err1}
  \varepsilon_l(x_i)
  =
  \frac{\sum_{j=0}^n w_j\,K(x_i, \sigma_{\rm I}, x_j) R_l(x_j) - \,L(x_i,\sigma_{\rm I})}{L(x_i,\sigma_{\rm I})}
}
after $l$ iterations and the mean square sum of
$\varepsilon_l(x_i)$  at every grid point

\eqn{
  \label{eqn:err2}
  {\cal E}_l
  =
  \frac{1}{n^2}\sum_{i=0}^n \left|\varepsilon_l(x_i)\right|^2
}
is used to check the quality of the iterated solution $R_l$ after $l$ recursions.
We remind here that this problem is ill-posed and therefore the solution is unstable
under perturbation of the input data. With increasing number of iterations
the solution will become more and more unstable. Therefore one has always to check 
whether the result $R_l$ is numerically still stable, {\it i.e.} free of unphysical 
oscillations.
Solving ill-posed problems requires a regularization scheme. This is naturally provided by 
this method through the grid $(x_i)$.

Let us remark at this point that 
the simple integration formula (eq.~(\ref{eqn:intcr})) has the advantage that all information 
entering the inversion process has the same weight except at the boundaries 
and it has constant grid gap, which provides a constant regularization 
over the whole integration region. 
In contrast, a Gauss-Legendre integration grid has 
smaller grid gaps at the boundaries than in the center of
the integration interval leading to a weaker regularization at the boundaries
and likely causing a lower quality of the inversion close to threshold. In addition
the integration kernel $K$ (eq.~(\ref{eqn:itlitK})) will 
become more easily numerically singular for the case that $L$ is given on a grid with many points,
{\it i.e.} more than about $50$ grid points. On the other hand
if $L$ is given on a sparse grid, the Gauss-Legendre integration is preferred, 
since it provides a more precise integration  for the same amount of grid points 
than formula (\ref{eqn:intcr}). For such a case the  mentioned disadvantages of the Gauss-Legendre
grid regarding regularization and numerical singularity are not so important any more.

The iteration (\ref{eqn:itlit}) is stopped if ${\cal E}_l$ is smaller 
than $2.5\cdot 10^{-16}$, if ${\cal E}_l$ or 
$\max_{i\in\{0,1,\dots,n\}}\{\varepsilon_l(x_i)\}$ is increasing or unphysical oscillations
arise  in the solution.
We allow at first some ``free'' iteration steps, depending on the quality of the input data, 
before we check if the iteration has to be terminated. This is necessary, since in 
the first iterations the errors ${\cal E}_l$ or $\max_{i\in\{0,1,\dots,n\}}\{\varepsilon_l(x_i)\}$
may increase temporarily due to large initial changes in $R_l$.

\subsection{Inversion using Banach's fix point theorem}

Let us first consider notations and conventions used in this subsection.
 We are working in the Hilbert space $L_2$ of the square integrable real
 functions
 $R(\omega) \equiv \braket{\omega}{R}$ with real arguments $\omega$ which fulfill
\begin{equation}\label{voll}
\int d\omega \,\ket{\omega}\bra{\omega} =1 \,\, .
\end{equation}
 The scalar
 product and  the norm  are  defined as usual:
\begin{equation}\label{scalar}
\braket{R_1}{R_2} := \int_{-\infty}^{\infty} d\omega \,
\braket{R_1}{\omega} \braket{\omega}{R_2} \equiv
\int_{-\infty}^{\infty} d\omega \, R_1(\omega) R_2(\omega)
\,\, ,
\end{equation}
and
\begin{equation}\label{norm}
||R|| := \sqrt{\braket{R}{R}}  < \infty  \,\,.
\end{equation}
For a fixed  $\sigma_{\rm I} > 0$, we are defining the Lorentz
 integration operator  ${\widehat L}(\sigma_{\rm I})$ (LIT) in accordance with
(\ref{eqLIT}) as follows:
\begin{equation}\label{lit1}
\bra{\omega}  {\widehat L}(\sigma_{\rm I}) \ket{R} \equiv
 L(\omega,\sigma_{\rm I})
 := \int d \omega'
\frac{1}{(\omega-\omega')^2 + \sigma_{\rm I}^2} R(\omega') \,\,, 
\end{equation}
where in general, if not noted  differently,
the upper and lower  limit in the integrals
 is  $\infty$ and $-\infty$, respectively.
 For the inversion 
method proposed in this subsection, let's introduce for a fixed 
Lorentzian  $L(\omega,\sigma_{\rm I})$ and a fixed $\sigma_{\rm I} > 0$ the following
 mapping  acting on an arbitrary function $R_1 \in L_2$: 

\begin{eqnarray}
\bra{\omega}{\hat T}(\sigma_{\rm I},L) \ket{R_1} &=:&
 {\hat T}(\sigma_{\rm I},L)(R_1(\omega)) \nonumber \\  &=& 
\frac{\sigma_{\rm I}}{\pi} L(\omega,\sigma_{\rm I})   - \frac{\sigma_{\rm I}}{\pi} 
\int d\omega' \frac{R_1(\omega')-R_1(\omega)}{(\omega'-\omega)^2 + \sigma^2_{\rm I}}  \,\, .\label{iter1}
\end{eqnarray}

As will be shown in Appendix \ref{app1}, this mapping has as fixpoint
just the desired response function $R(\omega)$ which creates via (\ref{lit1})
 the input Lorentzian $L(\omega,\sigma_{\rm I})$:
\begin{eqnarray}
{\hat T}(\sigma_{\rm I},L)(R(\omega)) &=& R(\omega) \quad \quad  \leftrightarrow
\nonumber \\
R(\omega) &=& 
\frac{\sigma_{\rm I}}{\pi} L(\omega,\sigma_{\rm I})   - \frac{\sigma_{\rm I}}{\pi} 
\int d\omega' \frac{R(\omega')-R(\omega)}{(\omega'-\omega)^2 + \sigma^2_{\rm I}} \,\, .  \label{fix1}
\end{eqnarray}

Now, in  appendix \ref{app2}, we will prove with the help of 
{\sc  Banach's Fixpoint Theorem}
that there exists an $\epsilon >
0$ so that for $0 < \sigma_{\rm I} < \epsilon$ 
the series ($n \geq 0$, $R^{(0)} \in L_2$ arbitrary)
\begin{equation}\label{iter11}
R^{(n+1)}(\omega) : = {\hat T}(\sigma_{\rm I},L)(R^{(n)}(\omega)) 
\end{equation}
 has a unique fix point, which is therefore just the searched
 response $R(\omega)$:

\begin{equation}\label{iter3}
R^{(n)}(\omega) \longrightarrow
 R^{\rm fix}(\omega) \equiv  R(\omega)
\quad \mbox{for} \quad n \rightarrow \infty \,\, ,  
\quad 0 < \sigma_{\rm I} < \epsilon \,\, .
\end{equation}

 This method  works therefore as follows: For a given Lorentzian
 $L(\omega,\sigma_{\rm I})$ with $\sigma_{\rm I}$ sufficiently small, 
 we choose some arbitrary starting function
 $R^{(0)}$ (in our practical applications, we have taken always 
 $R^{(0)}=0$ for the sake of simplicity) and then calculate
 the series (\ref{iter11}) till we have reached convergence. The found 
 fixpoint function $R^{\rm fix}$
 is then identical with the desired response function
 $R$. This method is therefore completely parameter free and does not
 need any suitable basic functions. 

In the version discussed so far, this approach has however one serious 
drawback. Due to the iterative procedure, it is obvious that the fixpoint
 $R^{\rm fix}$ is a $C^{\infty}$ function, {\it i.e.} 
 differentiable up to infinite order.
 In other words, this method is strictly valid only for response 
functions which   are $C^{\infty}$.   The
physical response function, however, is not analytic at the threshold energy
 $\omega_{\rm th}$ of the considered inclusive reaction and is therefore
 not $C^{\infty}$.
In order to get rid of this problem, one has to modify slightly the above
discussed algorithm. Instead of the range $]-\infty, \infty[$, all
occurring functions $L, R,...$ are now only defined on the interval
$[a,b]$, where $a\equiv \omega_{\rm th}$. The
 upper value $b$ should be in principle $\infty$. However, 
in practical applications, the Lorentzian $L$, which
 serves as a input in the algorithm, 
is known only in some finite range of $\omega$,
so that in practice $b$ is not infinite, but still ``large''. 

Instead of (\ref{lit1}) we have therefore to consider 
\begin{equation}\label{lit2}
 L_{(a,b)}(\omega,\sigma_{\rm I})
 := \int_{a}^{b} d \omega'
\frac{1}{(\omega-\omega')^2 + \sigma_{\rm I}^2} R(\omega') \,\, 
\end{equation}
 and the  counterpart of the linear mapping (\ref{iter1}) for finite
integrals becomes then
\begin{eqnarray}
 {\hat T}_{(a,b)}(\sigma_{\rm I},L)(R_1(\omega)) &=&
\frac{\sigma_{\rm I}}{\alpha_{(a,b)}(\omega)}
 L_{(a,b)} (\omega,\sigma_{\rm I})  \nonumber \\  & &
- \frac{\sigma_{\rm I}}{\alpha_{(a,b)}(\omega)}  \int_{a}^{b}
 d\omega' \frac{R_1(\omega')-R_1(\omega)}{(\omega'-\omega)^2 + \sigma^2_{\rm I}}  
\label{iter2} 
\end{eqnarray}
with the function
\begin{equation}\label{alpha}
\alpha_{(a,b)}(\omega) = \arctan(\frac{\omega-a}{\sigma_{\rm I}}) +
\arctan(\frac{b-\omega}{\sigma_{\rm I}}) \,\, .
\end{equation}

The series
\begin{equation}\label{iter22}
R_{(a,b) }^{(n+1)}(\omega) : = {\hat T}_{(a,b)}(\sigma_{\rm I},
L_{(a,b)}^R)(R_{ (a,b) }^{(n)}(\omega)) 
\end{equation}
has again, as will be shown in appendix \ref{app3}, the desired response
function $R$ as unique fixpoint for sufficiently small 
 $\sigma_{\rm I}$. So the method works also for functions which are defined
only on a finite range.

Summarizing, we have proven that this method allows for not too large $\sigma_{\rm I}$ a unique
determination of the  response function $R$ for a given  Lorentzian $L$.
As discussed in section \ref{results} in detail, it may however happen in practice that
$L$ is not very precisely known, containing for example oscillations due
to numerical uncertainties or approximations in its explicit calculation.
Therefore, similar to the Fridman approach,
 also in this method a sort of regularization has to be used
in order to avoid resulting unphysical oscillations in $R$. In practice, two
numerical tools are used which mutually supplement each other,
leading both to a reduction of the numerical resolution.
 First
of all, the number of mesh points in the occurring integrals can be reduced (in analogy 
to the Fridman method)
and second, the total number of iterations should be restricted, for example
  by choosing a suitable termination condition for ${\cal E}_l$ (\ref{eqn:err2}).

\section{Discussion of results}\label{results}

In the following we apply the various inversion methods of section~III to two 
different cases. First we investigate the efficiency of the regularization 
procedures implemented in the various inversion methods. As test case we 
consider a realistic example, where $L(\sigma_{\rm R},\sigma_{\rm I})$ is calculated from 
eqs.~(\ref{DE_incl},\ref{Lnorm}) without any knowledge of $R(\omega)$.
To this end we take $L(\sigma_{\rm R},\sigma_{\rm I})$ from ref.~\cite{LIT}, where the 
longitudinal response function for inclusive electron scattering off the 
deuteron at a momentum transfer of $q^2=5$ fm$^{-2}$ is considered. As described 
in ref.~\cite{LIT} eq.~(\ref{DE_incl}) was solved in an approximate
way thus leading to an $L(\sigma_{\rm R},\sigma_{\rm I})$ with a non-negligible error
margin. It is evident that for such a case a regularization must be present in
the inversion method otherwise reasonable results cannot be obtained.
Our second example serves for a different test, namely to check the ability
of the various inversion methods to precisely reproduce response functions with 
rather complicated structures. For this high precision check we take a test 
case, where the response function $R(\omega)$ is known beforehand analytically, and choose
a function with a double peak structure. Then $L(\sigma_{\rm R},\sigma_{\rm I})$ is 
calculated via eq.~(\ref{eqLIT}) with a rather high numerical precision and, 
finally, the obtained LIT is inverted.

\begin{figure}[ht]
\epsfig{file=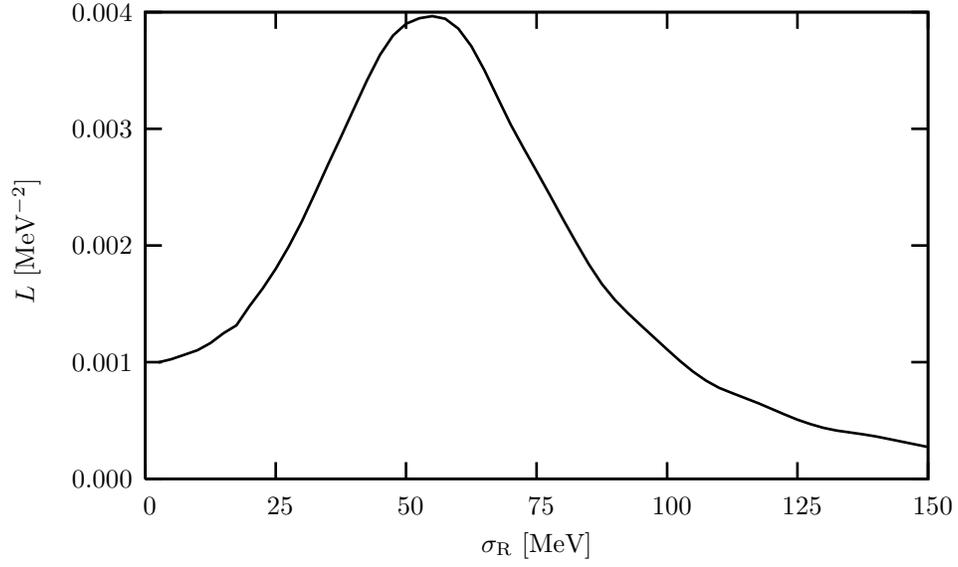}
\caption{\label{fig:02}
  LIT of the first test case taken from \cite{LIT}}
\end{figure}

The LIT of our first case is illustrated in fig.~\ref{fig:02}. One sees a pronounced
peak at about 50 MeV and one notes slight oscillations of the transform
at higher energies. As already mentioned the LIT is taken from \cite{LIT},
where the longitudinal response function $R_L(\omega,q)$ for inclusive
electron scattering off the deuteron at a momentum transfer of $q^2=5$ fm$^{-2}$
is considered. In ref.~\cite{LIT} eq.~(\ref{DE_incl}) was solved with
intention only approximately to obtain a transform with some error margin in 
order to work with a test case of not too high numerical precision.
For example the visible high-energy oscillations of the LIT in fig.~\ref{fig:02} are due to 
computational approximations. The solution of (\ref{DE_incl}) was obtained in \cite{LIT} 
by putting $\tilde\Psi(r)$ equal to zero at a neutron-proton distance of $r=20$ 
fm instead of using a more exact asymptotic behaviour. Of course in case of 
the deuteron a more precise LIT could be calculated easily.

We would like to emphasize once again that in case of such a
not very precisely determined LIT a proper regularization must be built into
the inversion method otherwise reasonable results cannot be obtained.
In \cite{LIT} the standard inversion method was used
for the inversion and problems were encountered, since the results
exhibited oscillations at lower energies. The problems arise from
Coulomb monopole (C0) and quadrupole (C2) transitions, which lead to a low-energy
shoulder in the response. As will be discussed also later on the standard
inversion method has difficulties to reproduce a more structured response
function. In \cite{LIT} these problems could be overcome by making separate
inversions for the C0 and C2 LITs.

\begin{figure}[ht]
\epsfig{file=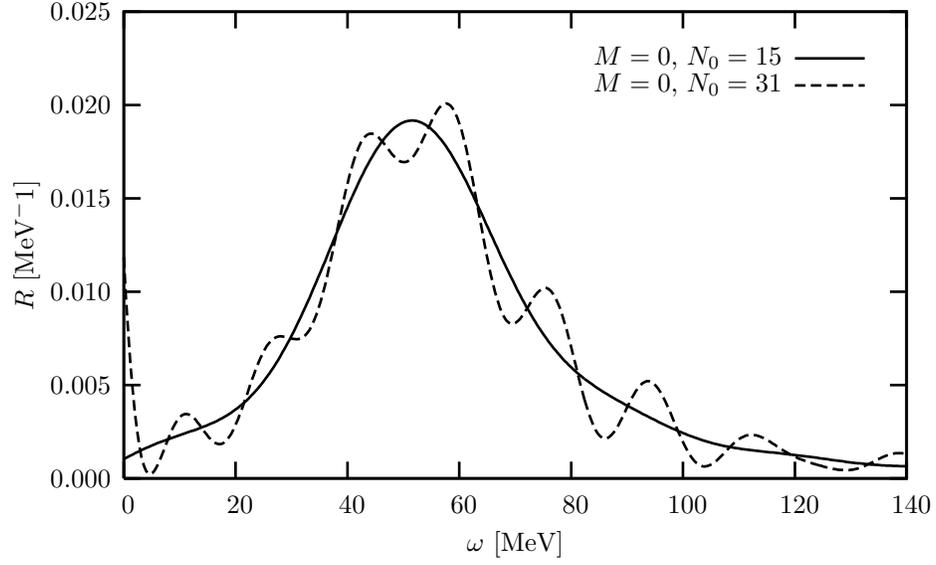}
\caption{\label{fig:03}
  Wavelet inversion results for the first test case
  with  $M=0$ (full: $N_0=15$, dashed: $N_0=31$).}
\end{figure}

First we discuss the inversion results with the wavelet method. Here we should
mention that $R_L$ contains also an elastic contribution. Thus it is necessary
to introduce an additional $\delta$-shape basis functions which accounts for
it. This can be achieved easily and the inversion 
result leads not only to the inelastic response function, but also to a value 
of the elastic form factor at $q^2=5$ fm$^{-2}$. In principle one could
also calculate the elastic form factor separately and then subtract its 
contribution from the LIT. In fig.~\ref{fig:03} we show two inversion results. While
the inversion with the parameters $M=0$ and $N_0=15$ leads to a smooth curve,
one already finds strong unrealistic oscillations if $N_0$ is increased to
31. Even stronger oscillations are found for the inversions with the other
parameter values of table~1. It shows that for the present example a rather 
low resolution, {\it i.e.} a rather strong regularization, is necessary in order to 
obtain an $R_L$ result which can be interpreted as realistic.

In contrast to the wavelet method the Banach inversion and the Fridman
method are iterative methods. The nature of the iterative inversion method
is that with increasing number of iterations the error ${\cal E}_l$ 
 (eq.(\ref{eqn:err2})) will become smaller and smaller but the solution will
become more and more unstable. That means that the amplitudes of unphysical
oscillations will increase. In the following the integration grid
is used as regularization. Therefore the larger the grid gaps are the smaller the
frequencies of the unphysical oscillations will be. In principle one could also 
introduce a regularization operator and  calculate the optimal
number of iterations depending on the desired error ${\cal E}_l$ 
 (eq.(\ref{eqn:err2})). We did not follow this approach of regularization
since the grid-regularization works successfully as discussed in the following.
To simplify the inversion process the elastic contribution to the LIT, 
$L_{\rm el}=0.0436/((E_0-\sigma_{\rm R})^2+\sigma_{\rm I}^2))$ with 
$-E_0=E_{\rm B}=2.2246\,\mbox{MeV}$ and $\sigma_{\rm I}=10\,\mbox{MeV}$ \cite{LIT}, 
has been removed in both inversion methods.

\begin{figure}[ht]
\epsfig{file=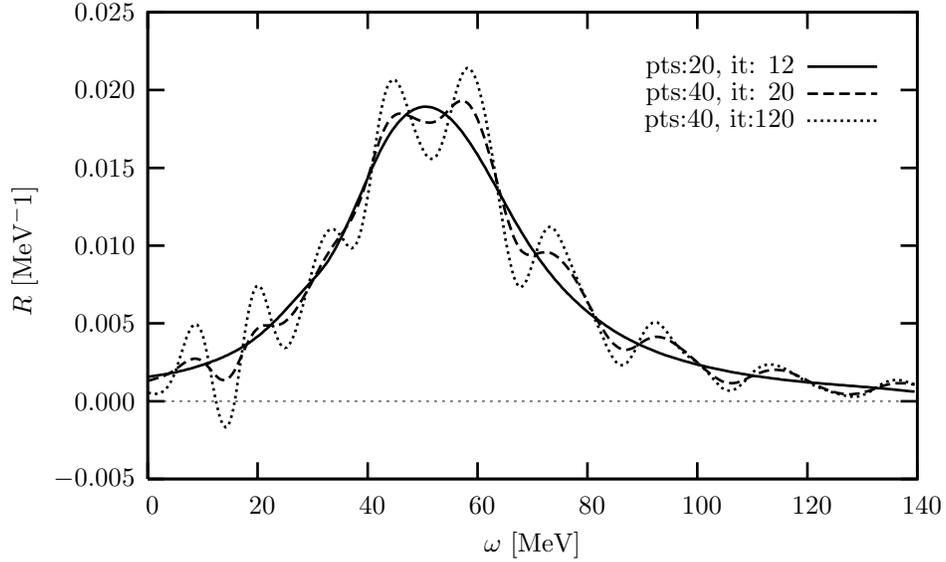}
\caption{\label{fig:04}
  Fridman inversion results for the first test case 
  (full: 20 grid-points and 12 iterations,
  dashed: 40 and 20,
  dotted: 40 and 120).}
\end{figure}

\begin{figure}[ht]
\epsfig{file=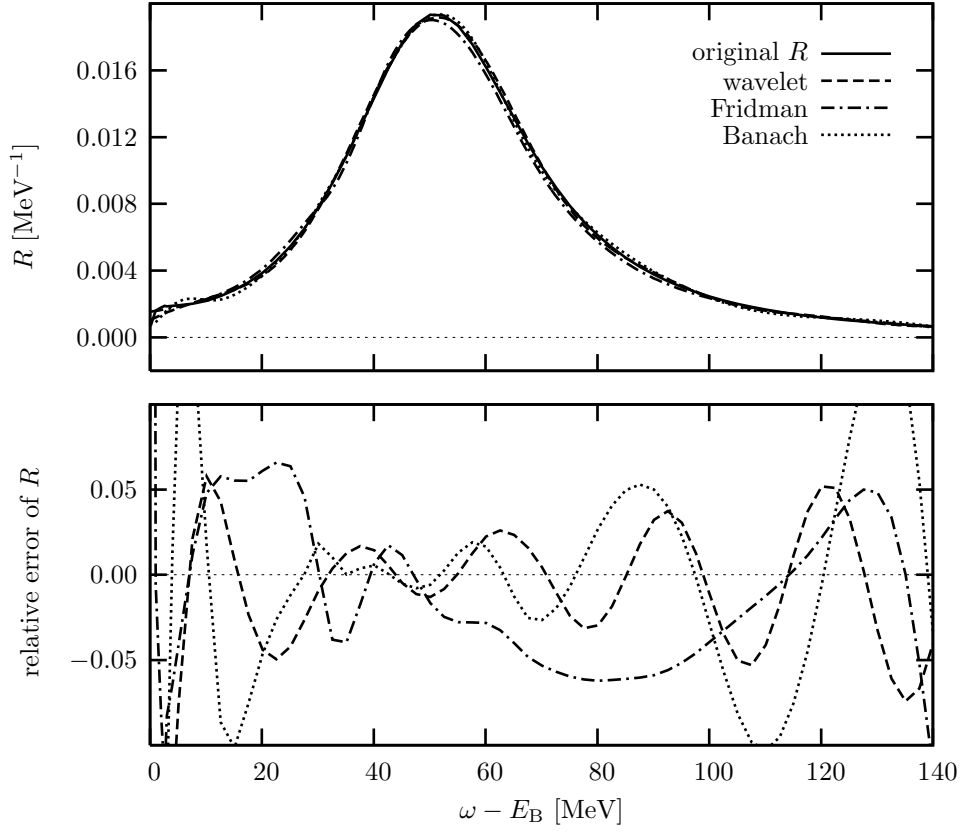}
\caption{\label{fig:05}
  Final inversion results with the three new  methods for the first test case
  are shown  (upper panel; full: original response function, dashed: wavelet, 
  dashed-dotted: Fridman, dotted: Banach) 
  along with their relative error (lower panel).}
\end{figure}

In fig.~\ref{fig:04} we illustrate examplarily the procedure 
to determine the inversion for the Fridman case.  First the number of iterations 
is fixed  to 120 and the integration is done on a grid with 40 points 
(curve: pts:40, it:120). This curve shows still quite large unphysical
oscillations although the integration grid consists of only 40 grid-points. By
further reducing the amount of iterations to 20 (curve: pts:40, it: 20) the
amplitude of the unphysical oscillations is reduced.
Then we lower the amount of grid points till a smooth curve is obtained. At that point
a strong regularization is  introduced. To improve the threshold region
the grid $(x_i)$ is split up into two grids, a small inner grid using the simple integration
method (eq.~(\ref{eqn:intcr})) and an outer grid using a Gauss-Legendre integration grid. 
Finally a systematic variation of the number of grid points, grid boundaries
 and the number of
iterations yields a  parameter set where the result is stable under small changes
of these parameters. The result shown fig.~\ref{fig:04} and fig.~\ref{fig:05} is obtained 
with 12 iteration where 6 points are used for the inner grid $[0,30]$ MeV and
14 points for the outer one   $]30,197]$ MeV.

A similar pattern occurs also for the Banach method. 
 Concerning the chosen resolution, we use  here 24 Gaussian mesh
points for the evaluation of the integrals and stop the iteration process
for ${\cal E}_l < 10^{-5}$. In contrast, if the Lorentzian is numerically
better  determined, like in our second example discussed below, typically
a couple of hundreds of mesh points and   termination conditions
 like        ${\cal E}_l < 10^{-15}$ (or even less) can be used  for small 
$\sigma_{\rm I}$  without
producing unphysical  numerical oscillations in the resulting response $R$.

Our deuteron example has the great advantage that $R(\omega)$ can also be 
determined directly from eq.~(\ref{Resp}), since two-nucleon continuum
wave functions can be calculated without problems. In fig.~\ref{fig:05} we compare
this response function, which is also taken from \cite{LIT}, to those 
from the three inversion methods.
It is evident that in spite of the strong regularization the inversion
results lead to an overall correct description and form a kind of 
error band. The wavelet and Banach inversions oscillate around the true
response with an error of less than 2 or 3 \% in the peak region. The
Fridman inversion exhibits a different pattern, but also describes the peak
region quite well. The relative errors become somewhat larger at lower and 
higher energies. Nonetheless in comparison to the standard inversion method,
which leads to very strong low-energy oscillations (see \cite{LIT}), 
our present results have a much smoother and more correct form at lower
energies. It is important to note that errors of the inversion results
are due to a not sufficiently precisely determined LIT. In fact as shown in
ref.~\cite{LIT} the error of the LIT amounts up to 1.5 \% in the peak region
and up to about 4 \% beyond 120 MeV. One sees that an error in the transform 
leads to an error of similar size in the inversion. Thus more precise
results could be obtained by calculating the LIT with a higher precision.
However, even with the present precision the low-energy shoulder could
be described much better if one makes separate inversions for the C0 and
C2 transition strength as was done in ref. \cite{LIT}.

\begin{figure}[ht]
\epsfig{file=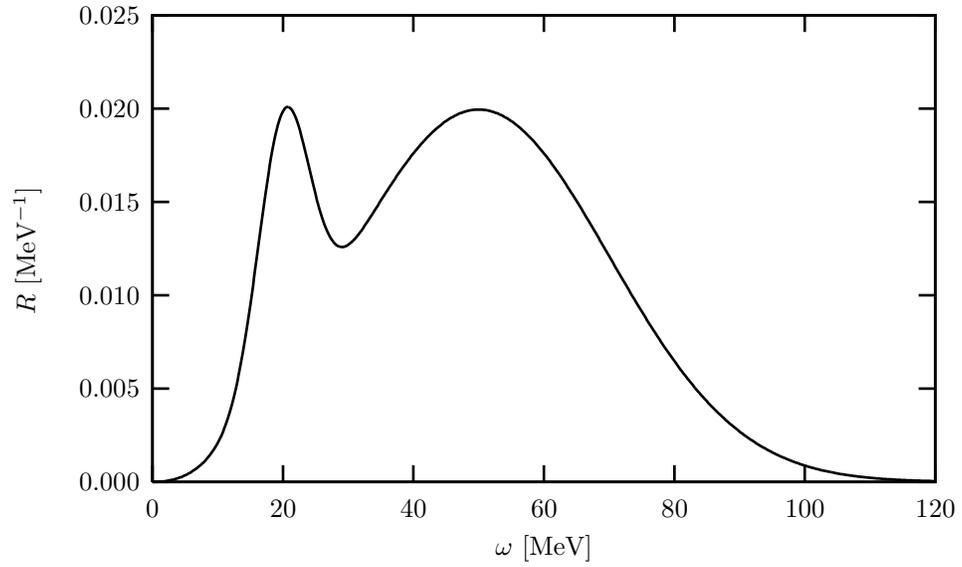}
\caption{\label{fig:06}
  Double peaked response function of second test
  case.}
\end{figure}

\begin{figure}[ht]
\epsfig{file=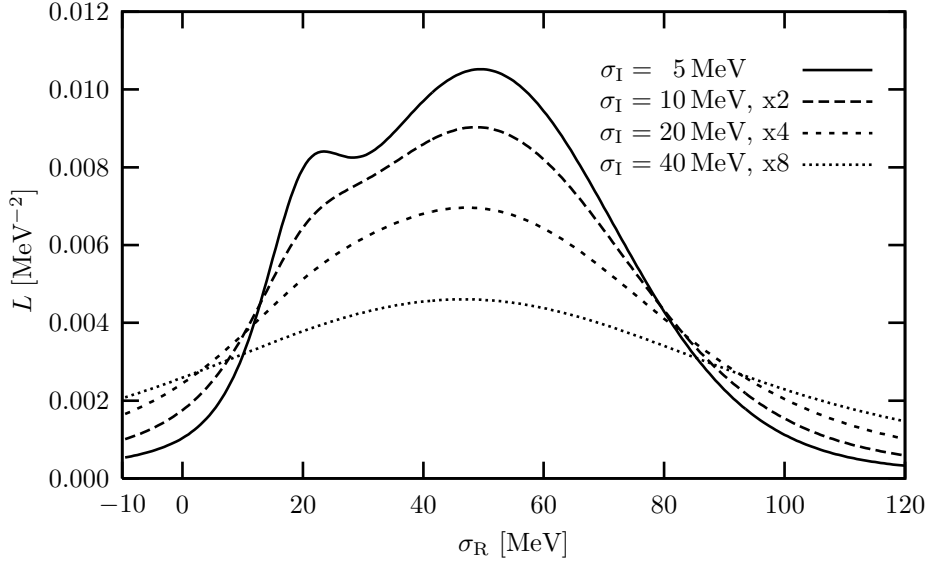}
\caption{\label{fig:07}
  LIT of the double peaked response function with different 
  $\sigma_{\rm I}$ (curves for $\sigma_{\rm I}=10$, 20 and 40 MeV
  are scaled by factors of 2, 4 and 8, respectively).}
\end{figure}

After our first example, which checked the proper implementation of the
regularization into the various inversion methods, we turn to
our high-precision test. In fig.~\ref{fig:06} we show the chosen $R(\omega)$  
up to $\omega=120$ MeV. The response function exhibits two peaks, one at 
$\omega=20$ MeV and the second at $\omega=50$ MeV. Both peaks have the 
same heights, but different widths. The corresponding LITs with $\sigma_{\rm I}=5$, 
10, 20, and 40 MeV are illustrated in fig.~\ref{fig:07}. It is evident that only 
the smallest value of $\sigma_{\rm I}$ leads also to two peaks for the 
transform, while the information about the two separate peaks is more 
and more smeared out with increasing $\sigma_{\rm I}$.

\begin{figure}[ht]
\epsfig{file=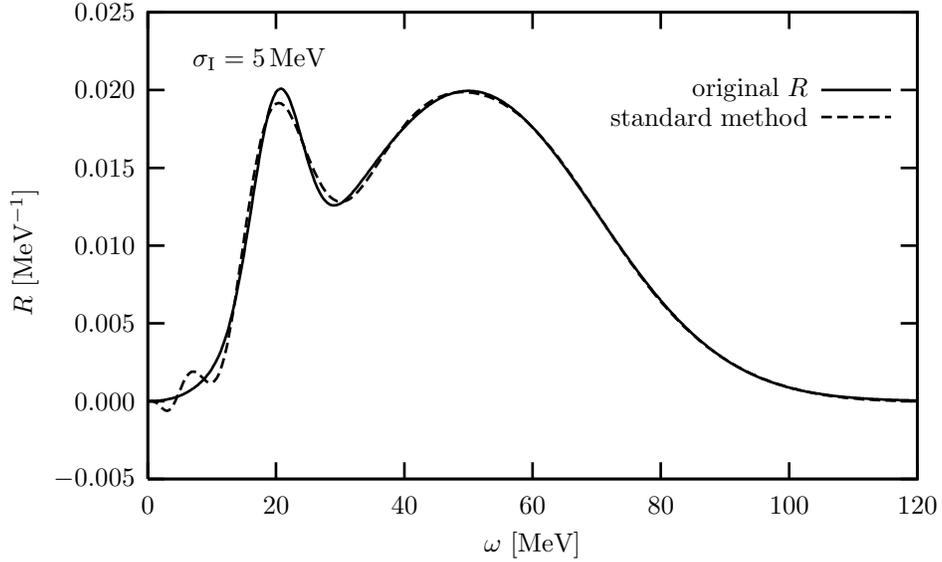}
\caption{\label{fig:08}
  Result for $\sigma_{\rm I}=5$ MeV (dashed) obtained with 
  the standard method (sec. IIIA) for the second  test case
  and the original double peaked response function (full).}
\end{figure}

In order to invert the LIT we calculate $L(\sigma_{\rm R},\sigma_{\rm I})$ for each of the 
four $\sigma_{\rm I}$ values from $\sigma_{\rm R}=-10$ MeV to $\sigma_{\rm R}=210$ MeV at 441 
equidistant grid points. It is instructive to use the standard LIT inversion 
method, described in section IIIA, for the transform with the  highest 
resolution ($\sigma_{\rm I}=5$ MeV). In fig.~\ref{fig:08} it is shown that the result is not 
of a very high quality. There are low-energy oscillations and the two peaks 
are not well reproduced. One sees that the standard inversion method has 
difficulties to reproduce a more structured response function like our present 
double peak example. The reason for the problem are the basis functions. They 
have a too large extension in the $\omega$-space, therefore oscillations are
introduced easily. In case of a more simpler structure like a single peak 
response such difficulties do not appear (see {\it e.g.} \cite{LIT}). We do not
show results with the standard inversion method for larger values of $\sigma_{\rm I}$.
As one may expect they lead to even worse results.

\begin{figure}[ht]
\epsfig{file=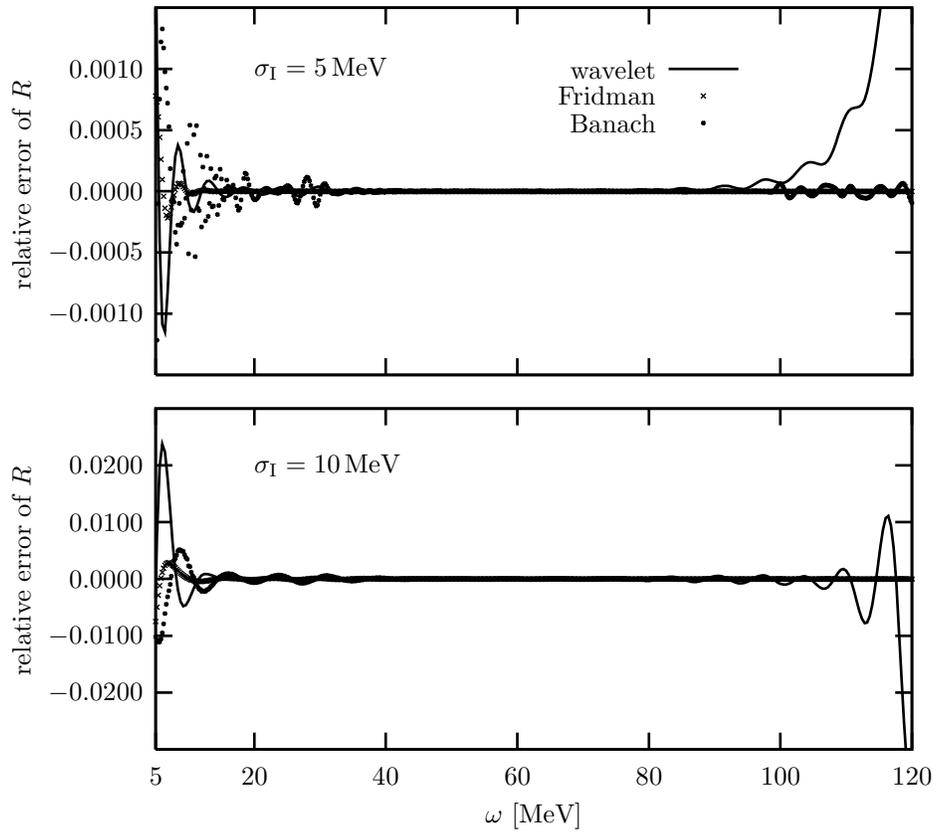}
\caption{\label{fig:09}
  Relative error of the inversion results for the second test case 
  obtained with the new methods 
  (full: wavelet, crosses: Fridman, points: Banach) 
  for $\sigma_{\rm I}=5$ MeV (upper panel)
  and for $\sigma_{\rm I}=10$ MeV (lower panel).
}
\end{figure}

\begin{figure}[ht]
\epsfig{file=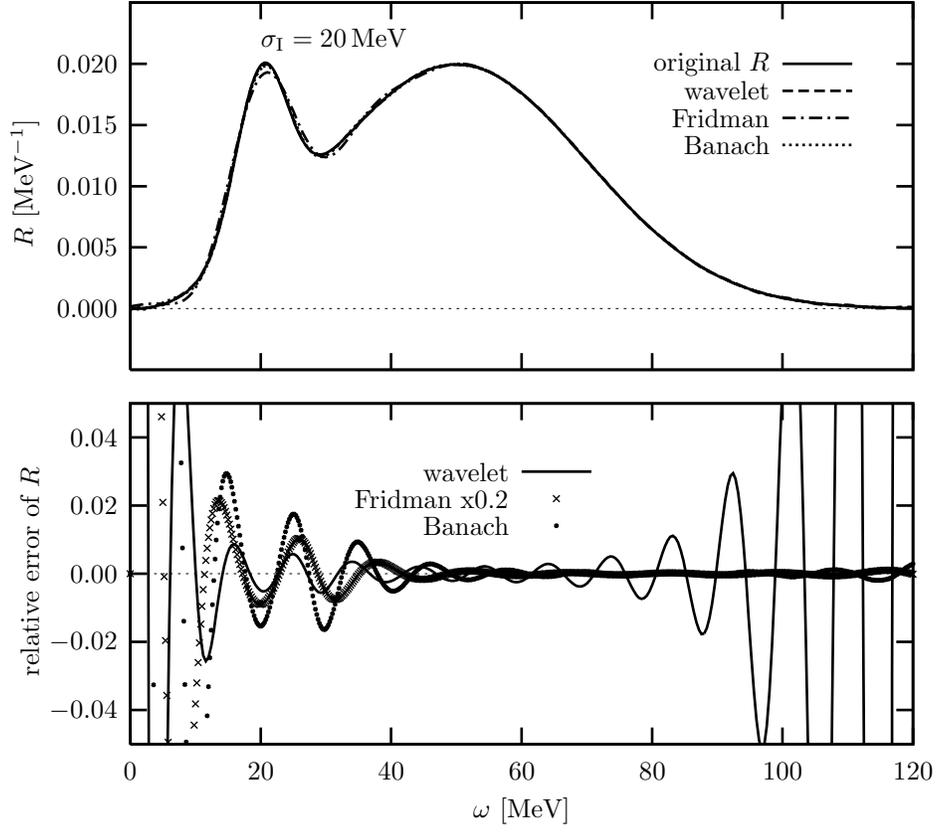}
\caption{\label{fig:10}
  Results of the three new inversion 
  methods  (full: original response function, dashed: wavelet, 
  dashed-dotted: Fridman, dotted: Banach) for 
  the second test case with $\sigma_{\rm I}=20$ MeV 
  are shown (upper panel) along with their relative error (lower panel;
  full: wavelet, crosses: Fridman scaled by a factor of $0.2$, 
  points: Banach).}
\end{figure}

\begin{figure}[ht]
\epsfig{file=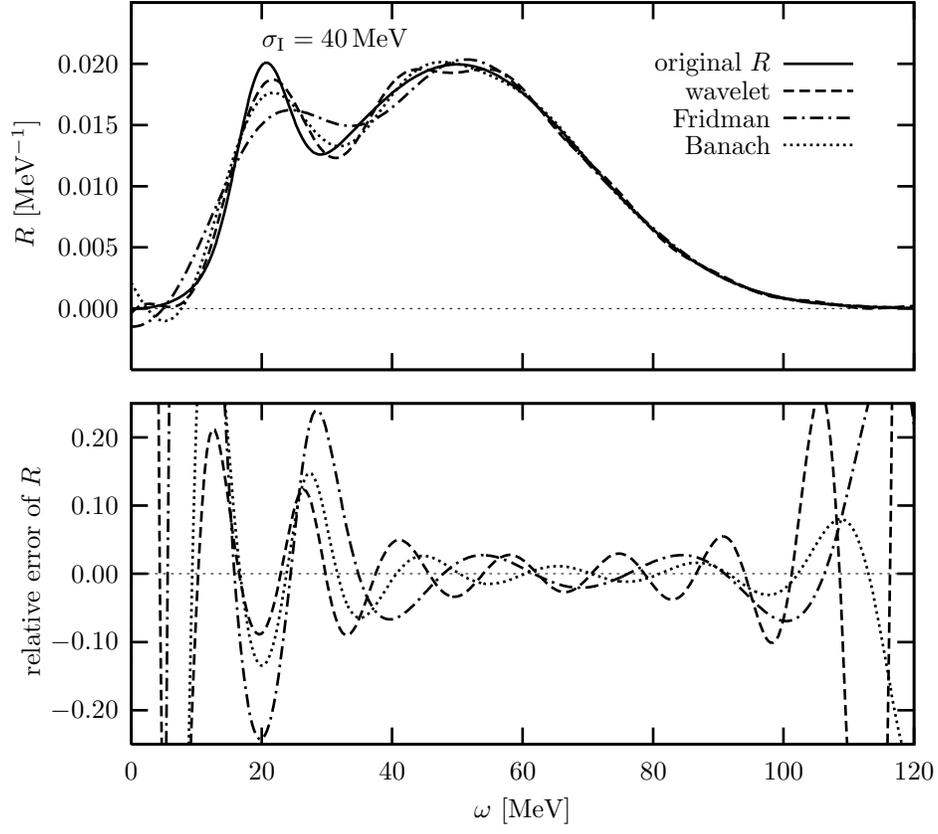}
\caption{\label{fig:11}
  Results of the three new inversion 
  methods (full: original response function, dashed: wavelet, 
  dashed-dotted: Fridman, dotted: Banach)
  for the second test case with $\sigma_{\rm I}=40$ MeV 
  are shown (upper panel) along with their relative error (lower panel).}
\end{figure}

It is obvious that alternative inversion methods are needed in case of more 
structured response functions. In fact the three inversion techniques described
in sections IIIB-D lead to much better results. For $\sigma_{\rm I}=5$ and 10 MeV 
one obtains results so close to the true $R$ that in fig.~\ref{fig:06} they would all appear 
identical to the true response function. Thus in fig.~\ref{fig:09} we show the 
relative errors of $R$ in an $\omega$-range, where $R(\omega)$ is not too close 
to 0. It is evident that the relative errors are very small: generally much less 
than 0.001 for $\sigma_{\rm I}=5$ MeV and less than 0.01 for $\sigma_{\rm I}=10$ MeV, only 
close to the low- and high-energy borders, where $R$ approaches 0 the error can 
become a bit larger. Here we would like to point out that the threshold region
can also be inverted with greater care (see final paragraph of subsection IIIA). 
In figs. \ref{fig:10} and \ref{fig:11} we illustrate the inversion results for 
the two higher $\sigma_{\rm I}$ values. One sees that one still obtains rather good 
results with $\sigma_{\rm I}=20$ MeV, particularly with wavelet and Banach's fix 
point theorem inversion techniques. With $\sigma_{\rm I}=40$ MeV the inversion results 
become worse. The shapes of the two peaks are not very well reproduced, but at 
least the peaks are recognized as two separate structures and also the 
high-energy tail is still rather well described. However, it is obvious that 
for the present case a $\sigma_{\rm I}$ value of 40 MeV is not sufficient to have 
precise inversion results. On the other hand with a further increase of the 
numerical precisions in calculating the transform and the inversions, which 
is nothing else than a further reduction of the regularization, one could have
also for $\sigma_{\rm I}=40$ MeV much better results.

\begin{figure}[ht]
\epsfig{file=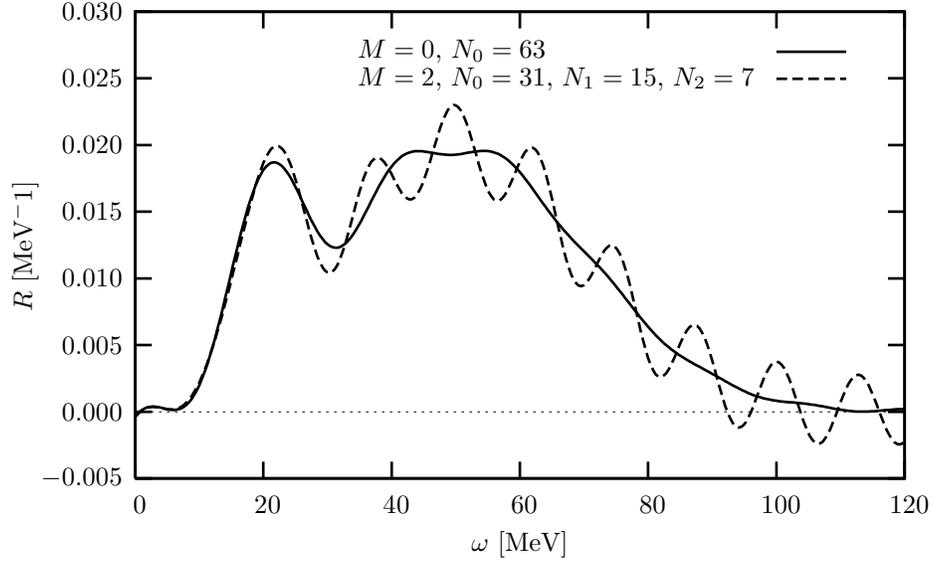}
\caption{\label{fig:12}
  Wavelet inversion results  
  for second test case
  (full: $M=0$, $N_0=31$, dashed: $M=2$, $N_2=31$, $N_1=15$, $N_2=7$).}
\end{figure}

In the above example Fridman and Banach methods lead to a single inversion
result for a given $\sigma_{\rm I}$, while the wavelet technique leads for any set 
of values for $M$ and $N_m$ (see table~I) to principally different inversion
results, in absence of unphysical oscillations they are of course
almost identical. For $\sigma_{\rm I}=5$, 10, and 20 MeV the result
with the best description of  $L(\sigma_{\rm R},\sigma_{\rm I})$ (sum of quadratic errors) 
is taken ($\sigma_{\rm I}=5$ and 10 MeV: $M=2$ and $N_0=N_1=N_2=63$, $\sigma_{\rm I}=20$ MeV:
$M=2$ and $N_0=63$, $N_1=31$, $N_2=15$). For $\sigma_{\rm I}=40$ MeV,
however, the inversion result with the smallest error shows strong and 
unrealistic oscillations (see fig.~\ref{fig:12}) and is thus discarded. Also the 14 next
best fits lead to similar unrealistic oscillations showing that also for
this case a rather strong regularization is necessary. Only the 16th best 
result ($M=0$, $N_0=63$) does not exhibit such an unrealistic pattern and
is therefore shown in fig.~\ref{fig:11}. Such strongly oscillating solutions like that
of fig.~\ref{fig:12} can be identified easily as unrealistic, since a true response
function has to be positive definite and in addition it is difficult to
imagine that a true response can exhibit such a regular oscillation pattern. 

We summarize our results as follows. The various newly introduced inversion
techniques for the Lorentz Integral Transform  contain a proper
regularization method and are thus able to treat the ill-posed
inversion problem. In addition all these techniques are capable to invert
also transforms of responses with rather complicated structures with
a very high precision. For such cases they lead to considerably better
inversion results than the standard LIT inversion method. Our results show
that the LIT approach is not restricted to treat cases, where simple
structures appear, but is suitable for many different applications.

\appendix

\section{Mathematical proofs concerning the fixpoint method}

\subsection{Proof of (\ref{fix1})}\label{app1}
In order to prove (\ref{fix1}), let's consider the expression
\begin{equation}\label{z1}
B(\omega,\sigma_{\rm I}) := \Im \int d\omega' \frac{R(\omega')}{\omega'-\omega - i \sigma_{\rm I}} \,\, ,
\end{equation}
which can be rewritten in two different manners: First, one has the identity
\begin{equation}\label{z2}
B(\omega,\sigma_{\rm I}) =   
\Im \int d\omega' \frac{R(\omega') (\omega'-\omega+i \sigma_{\rm I}) }{(\omega'-\omega)^2 + \sigma_{\rm I}^2}
= \sigma_{\rm I} L(\omega,\sigma_{\rm I})
\end{equation}
and second, one can write 
\begin{equation}\label{z3}
 B(\omega,\sigma_{\rm I}) =  
\Im \int d\omega' \frac{R(\omega')-R(\omega)}{\omega'-\omega - i \sigma_{\rm I}}
+
\Im \left(R(\omega) \int d\omega' \frac{1}{\omega'-\omega - i \sigma_{\rm I}}\right) \, .
\end{equation}
The second integral  on the right hand part of (\ref{z3}) can be easily solved
with the help of 
\begin{equation}\label{z31}
\frac{a}{\pi} \int_{-\infty}^{\infty} dx \frac{1}{x^2+a^2} = 1 \,\, ,\quad
 \quad (a  > 0) \,\, .
\end{equation}
Comparison of (\ref{z3}) and (\ref{z2}) yields  then (\ref{fix1}).

\subsection{Proof of uniqueness of iteration procedure}\label{app2}
In this appendix we prove that there exists  an $\epsilon > 0$ so
that  for all $\sigma_{\rm I}  < \epsilon$ the series (\ref{iter11})
converges uniquely against a function $R^{\rm fix}$ which is identical
with the desired response $R(\omega)$.

For that purpose, let us introduce the difference 
\begin{equation}\label{bew3}
\Delta^{(n)} : = R - R^{(n)} \, .
\end{equation}
 Exploiting (\ref{iter1}),(\ref{fix1}),(\ref{iter11}), the 
 function $\Delta$ fulfills 
\begin{equation}\label{linear1}
\Delta^{(n+1)} : = {\cal T}\Delta^{(n)}
\end{equation}
with the linear mapping
\begin{equation}\label{bew4}
{\cal T}\Delta^{(n)}(\omega)
 =    - \frac{\sigma_{\rm I}}{\pi} 
\int d\omega' \frac{\Delta^{(n)}(\omega')-\Delta^{(n)}(\omega)}{(\omega'-\omega)^2 + \sigma^2_{\rm I}} \, . 
\end{equation}

At next, we will prove that $\cal{T}$ is  
-- for sufficiently small $\sigma_{\rm I}$ -- a 
 contractive
 linear mapping, {\it i.e.}
there exists a constant $0 < q < 1$ so that 
\begin{equation}\label{bew41} 
   ||{\cal T}\Delta^{(n)}||^2 \equiv    ||\Delta^{(n+1)}||^2
\leq q  ||\Delta^{(n)}||^2 \,\, .
\end{equation}

{\bf Proof:}

For the norm of $\Delta^{(n+1)}$ we obtain
\begin{eqnarray}
  ||\Delta^{(n+1)}||^2 &=& \int d\omega  \Delta^{(n+1)}(\omega) 
\Delta^{(n+1)}(\omega) \nonumber \\
 & &
 \!\!\!\!\!\!\!\!\!\!\!\!\!\!\!\!\!\!\!\!\!\!\!\!\!\!\!\!\!
 \!\!\!\!\!\!\!\!\!\!\!\!\!\!\!\!\!\!\!\!\!\!\!\!\!\!\!\!\!
=  \frac{\sigma_{\rm I}^2}{\pi^2} \int d\omega \int d\omega' \int d\omega''
 \left( \frac{\Delta^{(n)}(\omega') - \Delta^{(n)}(\omega)}{(\omega' - \omega)^2 
+ \sigma_{\rm I}^2}\right) 
 \left( \frac{\Delta^{(n)}(\omega'') - \Delta^{(n)}(\omega)}{(\omega'' - \omega)^2 
+ \sigma_{\rm I}^2}\right) \,\, ,\label{bew5}
\end{eqnarray}
which can be rewritten as 
\begin{eqnarray}
  ||\Delta^{(n+1)}||^2 &=&
  \frac{\sigma_{\rm I}^2}{\pi^2} \int d\omega \int d\omega' \int d\omega''
 \left( \frac{1}{(\omega' - \omega)^2 
+ \sigma_{\rm I}^2}\,\, \right) \,\, 
  \left( \frac{1}{(\omega''- \omega)^2 
+ \sigma_{\rm I}^2}\right) \nonumber \\ & & 
\!\!\!\!\!\!\!\!\!\!\!\!\!\!\!\!\!\!\!\!\!\!\!\!\!\!\!\!\!\!
\!\!\!\!\!\!\!\!\!\!\!\!\!\!\!\!
 \left(\Delta^{(n)}(\omega') \Delta^{(n)}(\omega'') +
\Delta^{(n)}(\omega) \Delta^{(n)}(\omega) -
\Delta^{(n)}(\omega') \Delta^{(n)}(\omega) -
\Delta^{(n)}(\omega'') \Delta^{(n)}(\omega) \right) \,\, .   \nonumber \\ \label{bew6a}
\end{eqnarray}
This expression can be simplified with the help of (\ref{z31}) and
with the 
identity (which can be easily proven using the method of residues)
\begin{equation}\label{z32} 
\int_{-\infty}^{\infty} d\omega 
 \left( \frac{1}{(\omega' - \omega)^2 
+ \sigma_{\rm I}^2}\,\, \right) \,\, 
  \left( \frac{1}{(\omega''- \omega)^2 
+ \sigma_{\rm I}^2}\right) = \frac{2\pi}{\sigma_{\rm I}} 
 \left( \frac{1}{(\omega' - \omega'')^2 
+ 4\, \sigma_{\rm I}^2}\,\, \right) 
\end{equation}
as follows:
\begin{equation}\label{bew6}
  ||\Delta^{(n+1)}||^2 =
  ||\Delta^{(n)}||^2 
+ \frac{2 \sigma_{\rm I}}{\pi}
\bra{\Delta^{(n)}}{\widehat L}(2 \sigma_{\rm I}) \ket{\Delta^{(n)}} 
- \frac{2 \sigma_{\rm I}}{\pi}
\bra{\Delta^{(n)}}{\widehat L}(\sigma_{\rm I}) \ket{\Delta^{(n)}} \,\, .
\end{equation}

For further exploitation, let's prove at next the following

{\bf Lemma:} There exists an $\epsilon  > 0$, so that for all
 $\sigma_{\rm I}  < \epsilon$ 
\begin{equation}\label{bew7}
\frac{2 \sigma_{\rm I}}{\pi}
 \bra{\Delta^{(n)}}{\widehat L}(2 \sigma_{\rm I}) \ket{\Delta^{(n)}} 
 \leq A(\sigma_{\rm I}) \braket{\Delta^{(n)}}{\Delta^{(n)}}, 
\end{equation}
where the upper limit $A(\sigma_{\rm I})$ for 
$\sigma_{\rm I} < \epsilon$ fulfills the estimate
\begin{equation}\label{bew8} 
1- \delta  \leq A \leq   1 + \delta, \quad \quad \delta  <
\frac{1}{3}\,\, .
\end{equation}

{\bf Proof of the Lemma:}

Due to 
\begin{equation}\label{z103}
\lim_{\sigma_{\rm I} \rightarrow \infty} \frac{\sigma_{\rm I}}{\pi} 
\frac{1}{(x-y)^2+\sigma_{\rm I}^2} = \delta(x-y)\,\, ,
\end{equation}
  the statement is obviously true for  $\sigma_{\rm I} \rightarrow 0$
 with  
\begin{equation}\label{bew9}
\lim_{\sigma_{\rm I} \rightarrow 0} A(\sigma_{\rm I}) = 1\,\,.
\end{equation}
Because the LIT-operator in (\ref{lit1}) is depending analytically on
the parameter $\sigma_{\rm I}$, the statement follows immediately.

With the help of the Lemma, one obtains for (\ref{bew6}) the
following estimate:

\begin{eqnarray}
  ||\Delta^{(n+1)}||^2
 &=&
 |  ||\Delta^{(n)}||^2 
+ \frac{2 \sigma_{\rm I}}{\pi}
\bra{\Delta^{(n)}}{\widehat L}(2 \sigma_{\rm I}) \ket{\Delta^{(n)}} 
- \frac{2 \sigma_{\rm I}}{\pi}
\bra{\Delta^{(n)}}{\widehat L}(\sigma_{\rm I}) \ket{\Delta^{(n)}} 
| \nonumber \\
&\leq &   | 1+ (1+\delta) - 2(1-\delta) |  \quad   ||\Delta^{(n)}||^2 
\nonumber \\
&=& 3 \delta  ||\Delta^{(n)}||^2  \,\, .\label{bew10}
\end{eqnarray}
Due to 
\begin{equation}\label{bew11}  
q := 3\delta   < 3 \frac{1}{3} =1\,\, , 
\end{equation}
the mapping  $\Delta^{(n)} \rightarrow  \Delta^{(n+1)}$ defined via 
 (\ref{linear1}) is therefore for $\sigma_{\rm I}  < \epsilon$ contracting.

 In consequence, we can now apply  {\sc Banach's Fixpoint Theorem}: It states   
 that the series  $\Delta^{(n)}$ converges uniquely 
 against a fixpoint $\Delta$. Obviously,  (\ref{linear1}) is fulfilled by 
 $\Delta^{(n)}=0$ for all $n$, so that therefore 
\begin{equation}\label{bew12}  
 \Delta^{(n)} \rightarrow  \Delta  = 0 \,\, 
\end{equation}
and therefore, due to (\ref{bew3}), $R^{(n)} \rightarrow R$ uniquely.

\subsection{Incorporation of correct threshold behaviour of the response 
 function}\label{app3}
If the lower and upper value of the integral (\ref{lit1}) are not any
longer  $\pm \infty$, but some finite values $a$ and $b$, the proofs 
presented in appendix \ref{app1} and \ref{app2}
 have to be solely repeated where  instead of the
expression (\ref{z31}) its counterpart for finite integrals
\begin{equation}\label{z38}
\frac{a}{\alpha_{(a,b)}(\omega)} \int_a^b dx \frac{1}{x^2+a^2} = 1 \quad
 \quad (a  > 0)
\end{equation}
has to be used (the function $\alpha$ is given in (\ref{alpha})).
Therefore, instead of (\ref{fix1}) one has the identity
\begin{eqnarray}\label{z100}
R(\omega) &=& 
\frac{\sigma_{\rm I}}{\alpha_{(a,b)}(\omega)} L(\omega,\sigma_{\rm I})   - \frac{\sigma_{\rm I}}{
\alpha_{(a,b)}(\omega)} 
\int_a^b d\omega' \frac{R(\omega')-R(\omega)}{(\omega'-\omega)^2 + \sigma^2_{\rm I}}  \,\, ,
\end{eqnarray}
which motivates therefore the mapping  ${\hat
T}_{(a,b)}(\sigma_{\rm I},L)$ according to (\ref{iter2}).  

The counterpart of the difference function (\ref{bew3}) now becomes
\begin{equation}\label{z101}
\Delta_{(a,b)}^{(n+1)}(\omega) := R -R^{(n)}_{(a,b)} 
 =    - \frac{\sigma_{\rm I}}{\alpha_{(a,b)}(\omega)} 
\int_a^b d\omega' \frac{\Delta_{(a,b)}^{(n)}(\omega')-
\Delta_{(a,b)}^{(n)}(\omega)}{(\omega'-\omega)^2 + \sigma^2_{\rm I}}  \,\, .
\end{equation}

The norm of this function 
\begin{eqnarray}
  & & ||\Delta_{(a,b)}^{(n+1)}||^2 = \int_a^b d\omega  \Delta_{(a,b)}^{(n+1)}(\omega) 
\Delta_{(a,b)}^{(n+1)}(\omega) \nonumber \\
& &=  \int_a^b d\omega \int_a^b d\omega' \int_a^b d\omega''
  \left(\frac{\sigma_{\rm I}}{\alpha_{(a,b)}(\omega)} \right)^2
 \left( \frac{1}{(\omega' - \omega)^2 
+ \sigma_{\rm I}^2}\,\, \right) \,\, 
  \left( \frac{1}{(\omega''- \omega)^2 
+ \sigma_{\rm I}^2}\right) \nonumber \\ & & 
 \left(\Delta_{(a,b)}^{(n)}(\omega') \Delta_{(a,b)}^{(n)}(\omega'') +
\Delta_{(a,b)}^{(n)}(\omega) \Delta_{(a,b)}^{(n)}(\omega) -
\Delta_{(a,b)}^{(n)}(\omega') \Delta_{(a,b)}^{(n)}(\omega) -\right. \nonumber\\
& & \left. 
\Delta_{(a,b)}^{(n)}(\omega'') \Delta_{(a,b)}^{(n)}(\omega) \right) \,\, .   \label{z102}
\end{eqnarray}
is much more complicated to be handled than its counterpart
 (\ref{bew6a}): Due to the
finite limits in the integrals, the method of residues 
 cannot any longer be used.
Moreover, the constant factor $\pi$ in (\ref{bew6a}) now turns into the
  complicated
 function $\alpha_{(a,b)}(\omega)$ so that a simple relation like (\ref{bew6}) 
is not any longer valid. However, recalling the representation (\ref{z103})
 of the $\delta$-function as well as   
(for arbitrary $a,b$; provided $a < \omega <  b$) 
\begin{equation}\label{z104}
\lim_{\sigma_{\rm I} \rightarrow 0} \alpha_{(a,b)}(\omega) = \pi \,\, , 
\end{equation}
it is obvious that for sufficiently small $\sigma_{\rm I}$ the influence of
the finite limits $a$ and $b$ in the integrals dies out so that again
the  mapping $\Delta_{(a,b)}^{(n)} \rightarrow \Delta_{(a,b)}^{(n+1)}$ is 
a contractive one.


\end{document}